# An approach to the author citation potential: Measures of scientific performance which are invariant across scientific fields

Pablo Dorta-González [a,*], María Isabel Dorta-González [b], Rafael Suárez-Vega [c]

[a] Universidad de Las Palmas de Gran Canaria, TiDES Research Institute, Campus de Tafira, 35017 Las Palmas de Gran Canaria, Spain. *E-mail*: pdorta@dmc.ulpgc.es
[b] Universidad de La Laguna, Departamento de Ingeniería Informática, Avenida Astrofísico Francisco Sánchez s/n, 38271 La Laguna, Spain. *E-mail*: isadorta@ull.es
[c] Universidad de Las Palmas de Gran Canaria, TiDES Research Institute, Campus de Tafira, 35017 Las Palmas de Gran Canaria, Spain. *E-mail*: rsuarez@dmc.ulpgc.es

* Corresponding author and proofs. E-mail: pdorta@dmc.ulpgc.es (P. Dorta-González).

## ABSTRACT

The citation potential is a measure of the probability of being cited. Obviously, it is different among fields of science, social science, and humanities because of systematic differences in publication and citation behaviour across disciplines. In the past, the citation potential was studied at journal level considering the average number of references in established groups of journals (for example, the crown indicator is based on the journal subject categories in the Web of Science database).

In this paper, some characterizations of the author's scientific research through three different research dimensions are proposed: production (journal papers), impact (journal citations), and reference (bibliographical sources). Then, we propose different measures of the citation potential for authors based on a proportion of these dimensions. An empirical application, in a set of 120 randomly selected highly productive authors from the CSIC Research Centre (Spain) in four subject areas, shows that the ratio between production and impact dimensions is a normalized measure of the citation potential at the level of individual authors. Moreover, this ratio reduces the between-group variance in relation to the within-group variance in a higher proportion than the rest of the indicators analysed. Furthermore, it is consistent with the type of journal impact indicator used. A possible application of this result is in the selection and promotion process within interdisciplinary institutions, since it allows comparisons of authors based on their particular scientific research.

*Keywords:* researcher assessment; author metric; bibliometric indicator; citation analysis; source normalization; citation potential.

**Highlights**

1. We provide some different characterizations of the research area at author level based on three dimensions: production (journal papers), impact (journal citations), and reference (bibliographical sources).
2. We propose some measures of the citation potential for authors, based on proportions between dimensions.
3. We compare the dimensions and proportions in a set of 120 randomly selected highly productive authors from the CSIC Research Centre (Spain) in four subject areas.
4. The ratio between production and impact dimensions reduces the between-group variance in relation to the within-group variance in a higher proportion than the rest of measures analysed. Furthermore, it is consistent with the type of journal impact indicator used.

## 1. Introduction

This work is related to author metrics and citation-based indicators for the assessment of researchers from a general bibliometric perspective. It is well known that in some scientific fields the average number of citations per publication (within a certain time period) is much higher than in other scientific fields. This is due to differences among fields in the average number of cited references per publication, the average age of cited references, and the degree to which references from other fields are cited. In addition, bibliographical databases such as the Web of Science and Scopus cover some fields more extensively than others (Moed, 2005).

For decades, the number of publications and the number of citations have been the two accepted indicators in ranking authors. Recently, alternative indicators which consider both production and impact have been proposed (Dorta-González & Dorta-González, 2011; Egghe, 2013). However, these indicators based on the h-index do not solve the problem when comparing authors from different fields of science. Different scientific fields have different citation practices and citation-based bibliometric indicators need to be normalized for such differences in order to allow for author comparisons. Given these large differences in citation practices, the development of bibliometric indicators that allow for between-field comparisons is clearly a critical issue (Waltman & Van Eck, 2013).

Many decisions with regard to the allocation of research funds and the assignment of positions are based on citation counts. However, it remains unclear whether citation-based indicators are the appropriate measures in judging a scientist's future quality (e.g., Mazloumian, 2012; García-Pérez, 2013; Penner et al., 2013a, 2013b).

The problem of field-specific differences in citation impact indicators comes from institutional research evaluation (Leydesdorff & Bornmann, 2011; Van Raan et al., 2010). This is because research institutes often have among their missions the objective of integrating interdisciplinary bodies of knowledge and they are generally populated by scholars with different disciplinary backgrounds (Leydesdorff & Rafols, 2011; Wagner et al., 2011).

There are statistical patterns, which are field-specific, that allow for the normalization of the impact indicators. Garfield (1979) proposes the term 'citation potential' for systematic differences among fields of science based on the average number of references per paper. For example, in the biomedical fields, long reference lists with more than fifty items are common, but in mathematics, short lists with less than twenty references are the standard (Dorta-González & Dorta-González, 2013a, 2013b). This variability is a consequence of the different citation cultures and can produce significant differences in citation-based indicators since the probability of being cited is affected. In this sense, the average number of references is the variable most used in the literature to justify the differences between fields of science, as well as the most employed in source-normalization (Leydesdorff & Bornmann, 2011; Moed, 2010; Zitt & Small, 2008). However, it is necessary to consider other sources of variance in the normalization process (Dorta-González & Dorta-González, 2013a, 2013c).

Traditionally, normalization of field differences has usually been based on a field classification system. In said approach, each publication belongs to one or more fields and the citation impact of a publication is calculated relative to the other publications in the same field. Most efforts to classify journals in terms of fields of science have focused on correlations between citation patterns (Leydesdorff, 2006; Rosvall & Bergstrom, 2008). An example of a field classification system is the *JCR subject category list* (Pudovkin & Garfield, 2002; Rafols & Leydesdorff, 2009). For these subject categories, Egghe & Rousseau (2002) propose the *aggregate impact factor,* taking all journals in a category as one meta-journal. Another example of a field classification system is the *Scopus subject areas*.

Nevertheless, the precise delineation between fields of science and the next-lower level specialties has until now remained an unsolved problem in bibliometrics because these delineations are fuzzy at any moment in time and develop dynamically over time. Therefore, classifying a dynamic system in terms of fixed categories can lead to error because the classification system is defined historically while the dynamics of science is evolutionary (Leydesdorff, 2012, p.359).

Recently, the idea of source normalization was introduced; which offers an alternative approach to normalizing field differences. In this approach, normalization is achieved by looking at the referencing behaviour of citing journals. Some indices, such as the *fractionally counted impact factor* (Leydesdorff & Bornmann, 2011; Zitt & Small, 2008) which divides each citation by the number of references, the *source normalized impact per paper –SNIP–* (Moed, 2010) which divides total citations by the median number of references in the citing journals, and the *topic normalized impact factor* (Dorta-González et al., 2014) which considers the aggregate impact factor of all citing journals, have been proposed.

In citation-based research evaluations, it is crucial to control the previously mentioned differences among fields. This is especially the case for performance evaluations at higher levels of aggregation, such as countries, universities, or multi-disciplinary research groups. The *crown indicator* (Van Raan et al., 2010) and the similar *normalized mean citation rate* (Glänzel et al., 2009) use a normalization mechanism that aims to correct the differences among fields. Given a set of publications, for each one, the number of citations it has received and its expected number of citations are determined. The expected number of citations of a publication equals the average number of citations of all documents (by any author) of the same type (i.e., article, letter, or review) published in the same field and in the same year in the database. To obtain the crown indicator, the sum of the actual number of citations of all publications and the sum of the expected number of citations of all publications are divided. An alternative mechanism (Lundberg, 2007; Opthof & Leydesdorff, 2010) is firstly to calculate, for each publication, the ratio of its actual number of citations and its expected number of citations and then take the average of the ratios obtained.

However, in these source-normalized metrics the expected number of citations is determined by the field which is defined in a field classification system. Therefore, these metrics do not include any great degree of normalization in relation to the specific research topic of each author. The topic normalization is necessary because different scientific topics have different citation practices. Therefore, citation-based bibliometric indicators need to be normalized for such differences between topics in order to allow for between-topic comparisons of authors. In this sense, we use the aggregate impact factor of three different sets of journals as a measure of the different dimensions in the citation potential of an author, and we employ a combination of these dimensions in the construction of a source normalized indicator to make it comparable between scientific fields. In order to test this new impact indicator, an empirical application with 120 authors belonging to four different fields is presented. The main conclusion we obtain is that our rate between production and impact dimensions reduces the between-group variance in relation to the within-group variance in a higher proportion than the rest of indicators analysed. Furthermore, it is consistent with the type of journal impact indicator used.

## 2. Dimensions and proportions of the author citation potential

Even within the same field, each researcher is working on one or several research lines that have specific characteristics, in most cases very distant from those of other researchers in the same field.

Generally, the citation potential in a field is determined within a predefined group of journals. This approach requires a classification scheme for assigning publications to fields. Given the fuzziness of disciplinary boundaries and the multidisciplinary character of many research topics, such a scheme will always involve some arbitrariness and will never be completely satisfactory. Therefore, we propose measuring the citation potential in the specific topic of each author and using this measure as an indicator of the probability of being cited in that topic.

The problem underlying the characterization of the author citation potential is as follows. Given a set of publications from an author in different journals and years, we will try to obtain a measure of the author topic defined by some dimensions of these publications so it can be compared with that of a different author (with publications in different journals and years). This problem arises in the evaluation of the research, when comparing authors from different research lines, especially authors with diverse backgrounds in research centres and institutes.

Let us consider a 5-year time window $Y$. In this paper, we propose characterizing the topic of an author in period $Y$ using three different dimensions: the weighted average of the impacts in the journals containing the author's papers in $Y$ (production dimension $P$), the weighted average of the impacts in the journals citing the author's papers in $Y$ (impact dimension $I$), and the weighted average of the impacts in the journals included as references in the author's papers in $Y$ (reference dimension $R$). The dimensions of the author citation potential are illustrated in Figure 1.

[Figure 1 about here]

In order to facilitate the reading of the paper, the notation used in the operational characterization of the author citation potential is shown in Table 1. In this characterization we propose the use of journal impact indicators instead of number of citations received by a particular paper. This is because it is necessary that several years pass after the publication of a document, so that the number of citations can be a consistent indicator in comparing similar documents of the same type published in the same year with that of other researchers in the same field. Consistency is a mathematical property based on the idea that the ranking of two units relative to each other should not change when both units make the same progress in terms of citations. Something similar happens when considering an indicator based on the percentage of highly cited publications, for example, the percentage of publications belonging to the top 5% or the top 10% of a particular field. In some fields (e.g., Economics) more than 5 years are needed to obtain a consistent measure of impact (Dorta-González & Dorta-González, 2013a). In many fields of the Humanities it is necessary to wait even longer.

[Table 1 about here]

However, in the evaluation of researchers for promotion and recruitment, the most recent production years of an author have a greater predictive power in their future production. Therefore, it is useful to know a measure of the author citation potential based on journal impacts in their topic.

We consider the following dimensions of the author's research area:
(*d1*) *The production dimension –P–* is the first measure of the probability of being cited in the research area and it is based on the author's publications. It is the weighted average of the impacts in the journals containing the author's papers in the target window. Therefore, this is the expected impact for the author.

As an example, the production dimension of A. Bocci (Physics & Astronomy) is illustrated in Table 2. Considering the journals in which Bocci's papers are published, the production dimension of this author is 2.817.

[Table 2 about here]

(*d2*) *The impact dimension –I–* is the second measure of the probability of being cited in the research area and it is based now on author citations. It is the weighted average of the impacts in the journals citing the author's papers in the target window. Therefore, this is the observed impact for the author's publications.

(*d3*) *The reference dimension –R–* is the third measure of the probability of being cited in the research area and it is based on author's bibliographical references in the target window. It is the weighted average of the impacts in the journals included as references in the author's papers in the target window.

In all three cases, the average is weighted by the number of papers in each journal, and the impact indicator of the journal corresponds to the year of publication. Through these different dimensions, the following four indicators that attempt to normalize the citation potential in the author's topic (dividing some dimensions by others) are proposed.

(*r1*) *The production over impact ratio –P/I–* is the proportion between production and impact dimensions. A quantity larger than one indicates that the author has published in journals with impact indicators above those observed for other authors in the same research area. This is because the average impact of the author's publications is compared with the average impact of the researchers citing this author. In this formulation only those publications in which the researchers cite this author are considered. Therefore, a value of 1.10 indicates that the production impact of the author is 10% higher than the other authors in the research area. Alternatively, a value of 0.80 indicates that the production impact of the author is 20% lower than the other authors in the research area.

As an example, in a similar way as in Table 2, the impact dimension of A. Bocci is 1.936, and therefore production over impact is 2.817 / 1.936 = 1.455. This quantity larger than one indicates that Bocci has published in journals with impact indicators higher than average in the same research topic. In particular, 1.455 indicates that the production impact of this author is 45% higher than other authors in the same research topic.

(*r2*) *The production over reference ratio –P/R–* is the proportion between the production and reference dimensions. (*r3*) *The impact over reference ratio –I/R–* is the proportion between the impact and reference dimensions. In both cases the interpretation is similar to the *P/I* case. Finally, (*r4*) *the production and impact over reference ratio –(P+I)/2R–* is the arithmetic mean between *P/R* and *I/R*, i.e., *(P/R + I/R)/2 = (P+I)/2R*.

A direct application of our methodology (an author citation potential obtained through journal impact indicators) is to identify those researchers who publish in higher impact journals than expected in their research topic. This would contextualize the topic of each author several years before knowing the real impact of their publications (through the received citations) in a consistent way.

In the empirical application we studied which of the previous ratios greater reduces the between-group variance in relation to the within-group variance in a set of 120 authors from four different fields.

## 3. Methods and materials

The bibliometric data was obtained from the online version of the Scopus database during the first week of April 2014. Only journal papers in the period 2009-2013 were included, considering for each journal two impact indicators (with a 3-year citation time window) in the year of its publication. The first indicator is the *scimago journal ranking –SJR–* (González-Pereira et al., 2009), which considers the prestige of the citing journals, and the second is the *source normalized impact per paper –SNIP–* (Moed, 2010), which divides total citations by the median number of references in the citing journals. The SNIP was used to compare results and identify the proper normalization and consistency of the ratios with the type of journal impact indicator employed.

Four subject areas were considered: Chemistry, Computer Science, Medicine, and Physics & Astronomy. This was motivated in order to obtain authors with systematic differences in publication and citation behaviour. We designed a random sample with a total of 120 authors (30 in each subject area). They were selected from the highly productive authors of the Consejo Superior de Investigaciones Científicas –CSIC– (Spain). In the population only those authors with a production over the mean in their subject area were considered.

We used seven indicators: three that measure different dimensions of the citation potential associated to the author, and four normalized indicators of the citation potential in the topic in which the author works.

## 4. Results and discussion

In the empirical application we studied which measure of the author citation potential produces a closer data distribution among subject areas in relation to its centrality and variability measures. We compared the seven indicators (three dimensions and four ratios) described in Table 1.

Table 3 shows the different dimensions and proportions of the author citation potential in the sample. Furthermore, three general production and impact indicators (number of papers, number of citations, and h-index) are shown. Table 3 presents two different scenarios, the first one (columns 6 to 12) considers the SJR as the impact indicator for journals, and the second one (columns 13 to 19) takes the SNIP as the impact indicator for journals. Thus, at any time the value of an indicator based on journal impacts using absolute citation frequencies can be compared to that based on relative citation frequencies.

[Table 3 about here]

In relation to the dimensions and proportions of the author citation potential (columns 6 through 19), important differences between both research areas and researchers within the same field can be seen in Table 3. This firstly reflects the peculiarities in the publication and citation habits of each research area as a whole and, secondly, the peculiarities in the specific research topic of each author. Furthermore, it can also be seen that, for each particular author, significant differences between the dimensions of the citation potential exist. The differences among the dimensions of the author citation potential are lower in the case of SNIP. This is expected because normalized impact indicators are used. Thus, in a major number of cases these differences are below 1 (SNIP), while in the case of SJR these differences are in many cases higher than 3.

For the general production and impact indicators (columns 3 to 5), note the variability in the data. In the case of the number of publications, the variation range is 402 (the difference between the maximum and the minimum value). Half of the authors have published less than 33 papers in the analysed period (the median is 33). However, the average number of publications per author is 66. Therefore, a small number of researchers have published the most papers (Bradford's Law). This is motivated by skewed distributions of data that are not centred on the mean. In the case of the number of citations, the variance is even larger, being 8,772 the variation range. Half of the authors have been cited less than 337 times in the period (the median is 337). However, the average citations per author is 1,058, again indicating that a small number of researchers have received most of the citations. Regarding the h-index, the range of variation is 41 and the median (10) is less than the mean (12.8).

Central-tendency and variability measures in the four subject areas are shown in Table 4. Note that for any dimension of the author citation potential, the values are very different from one research area to another. Notice the high differences between areas in medians, means, and standard deviations. This is because the subject areas considered in the sample are very different in relation to the publication and citation behavior. Furthermore, the medians are well below the means, indicating skewed distributions with many authors having low values and only a small number of authors with high values. However, when using ratios these differences between areas are greatly reduced.

[Table 4 about here]

Box-plots comparing the subject areas are shown in Figure 2. This is a way of graphically depicting the data in the sample through their quartiles. The left and right boundaries of the box are the first and third quartiles, and the band inside the box is the second quartile (the median). Therefore, the spacings between the different parts of the box indicate the degree of dispersion and skewness in the data. The ends of the whiskers represent the minimum and maximum of all of the data in the sample.

[Figure 2 about here]

As shown in the first row of Figure 2, there are large differences between subject areas. However, there is a certain pattern in the data distribution. Thus, for all three indicators, smaller values are observed in Computer Science, medium values are obtained in Chemistry and Medicine, and the highest values are observed in Physics & Astronomy. A similar behavior is observed both in *P* and *I* in the second row of Figure 2. However, both indicators produce fairly similar distributions of data between areas when normalized data is used (see SNIP in row 3), which does not occur in the case of *R*. Finally, with respect to the ratios in rows 4 to 6, the indicator that produces closer distributions of data between subject areas is *P/I* (with the exception of Physics & Astronomy). The differential behaviour of Physics & Astronomy is also observed when using normalized impact indicators (SNIP) and it is justified because in this subject area clearly higher values in the numerator of the ratio converge. In conclusion, *P/I* is the ratio based on non-normalized journal impacts that produces the least differences between most areas, and it is also close to the results using normalized journal impacts (SNIP). This suggests the possibility of focusing exclusively on non-normalized impact indicators because *P/I* is consistent with respect to the type of journal impact indicator.

The range of data variation in *P/I* is greater in Computer Science than in the other aforementioned areas. Notice the values are around one except in Physics & Astronomy where they are considerably above this value, and with a lower variability than the rest. Therefore, most authors in Physics & Astronomy have published in journals with an average impact indicator higher than the citing journals average impact indicators in their research topics (*P/I* is greater than one). In all other areas the median is about one.

Figure 3 compares the different dimensions of the author citation potential in the four subject areas. Authors have been listed by the impact dimension order. Overall, impact dimension is quite different from the other dimensions; so these three dimensions seem to offer different facets of the author's research.

[Figure 3 about here]

Now, we will test which normalization (ratio between dimensions) of the citation potential reduces the between-group variability in relation to the within-group variability. The central-tendency and the variability measures for the different dimensions and proportions of the author citation potential in the aggregate data are shown in Table 5. Moreover, it shows the within- and the between-group variability.

[Table 5 about here]

Within- and between-group variability are both components of the total variability in the combined distributions. What we are doing when we compute within- and between-group variability is to partition the total variability into the within and between components. So: within variability + between variability = total variability.

But, how do we measure variability in a distribution? That is, how do we measure how different scores are in the distribution from one another? In this work we use variance as a measure of variability. Recall that variance is the average square deviation of scores about the mean.

Note in Table 5 that the proportion between production and impact dimensions produces the greatest percentage reduction of the variance (76.3%). Using SNIP only an additional 3.9% reduction is obtained, but this improvement does not justify the use of this type of normalized impact indicator. In addition, the central-tendency measures of the data distributions are quite close to the SNIP scenario. The deviations are only 2.2% for the median and 3.1% for the mean.

Tables 6 and 7 provide the Pearson correlations and the Spearman rank correlations between different indicators of the author's research. A perfect Spearman correlation results when two indicators are related by any monotonic function. However, it contrasts with the Pearson correlation, which only gives a perfect correlation when the two measures are connected by a linear function. In this sense, the Spearman correlation is less sensitive than the Pearson correlation to strong outliers that are in the tails of both distributions. This is because Spearman coefficients limit the outlier to the value of its rank. However, in general, similar results are obtained in Tables 6 and 7.

[Tables 6 and 7 about here]

The general pattern that can be observed in the correlations reported in Table 6 is that the bibliometric variables (papers, cites, and $h$) are largely correlated, with all of the Pearson correlations above 0.54, and most of them above 0.80. However, in general, these variables are not correlated with $P$, $I$, $R$, and $P/I$, and therefore these measures seem to offer different facets of the author's research.

Between the number of papers and the number of citations, coefficients are greater than 0.92 in half of the areas. In the case of Physics & Astronomy, the productivity can explain more than 0.98% of the impact variance ($0.99^2 = 0.98$). The h-index correlates more with the number of citations (above 0.94 in half of the areas) than with the number of papers (above 0.83 in half of them). In general, the highest correlations are achieved in Physics & Astronomy, followed by Medicine and Chemistry. The lowest correlations between these variables are reached in Computer Science.
Note in Figure 4 distinct patterns in each subject area for the case of SJR and a more common bivariate distribution across subject areas for the case of SNIP. There are large correlations between the production and the impact dimensions of the research topic (above 0.80, except in Computer Science where it is 0.46). Therefore, in general publishing in higher impact journals improves visibility and produces citations in greater impact journals.

[Figure 4 about here]

Finally, for the *P/I* ratio the Pearson correlations between SJR and SNIP are large, in all the areas above 0.84 and in half of them above 0.91 (0.84 Chemistry; 0.89 Computer Science; 0.91 Medicine; 0.92 Physics & Astronomy; all of them significant at the 99% level). This indicates that the *P/I* ratio is consistent to the type of journal impact indicator.

## 5. Conclusions

Different scientific fields have different citation practices, and citation-based bibliometric indicators need to be normalized for such differences between fields in order to allow for between-fields comparisons of citation indicators. In this paper, we provide a normalization approach based on the dimensions of the author's research.

An empirical application, with 120 authors from four different subject areas, shows that the ratio between production and impact dimensions reduces the between-group variance in relation to the within-group variance in a higher proportion than the rest of the indicators analyzed in this paper. Furthermore, this normalized indicator is consistent in the sense that it is independent of the type of journal impact indicator considered.

The subject areas considered are very different in relation to the citation behavior. For this reason, in the sample there are important differences among the dimensions of the citation potential from one author to another. However, the proportion between production and impact dimensions is very close in all the subject areas considered.

We have developed a measure of scientific performance whose distributional characteristics are invariant across scientific fields. Such a measure would allow direct comparisons of scientists in different fields and permit a ranking of researchers that is not affected by differential publication and citation practices across fields.

Finally, it is necessary to be cautious when comparing authors from different subject areas. Many decisions with regard to the allocation of research funds and the assignment of positions are based on citation counts. However, it remains unclear whether citation-based indicators are appropriate measures to judge a scientist's future quality.

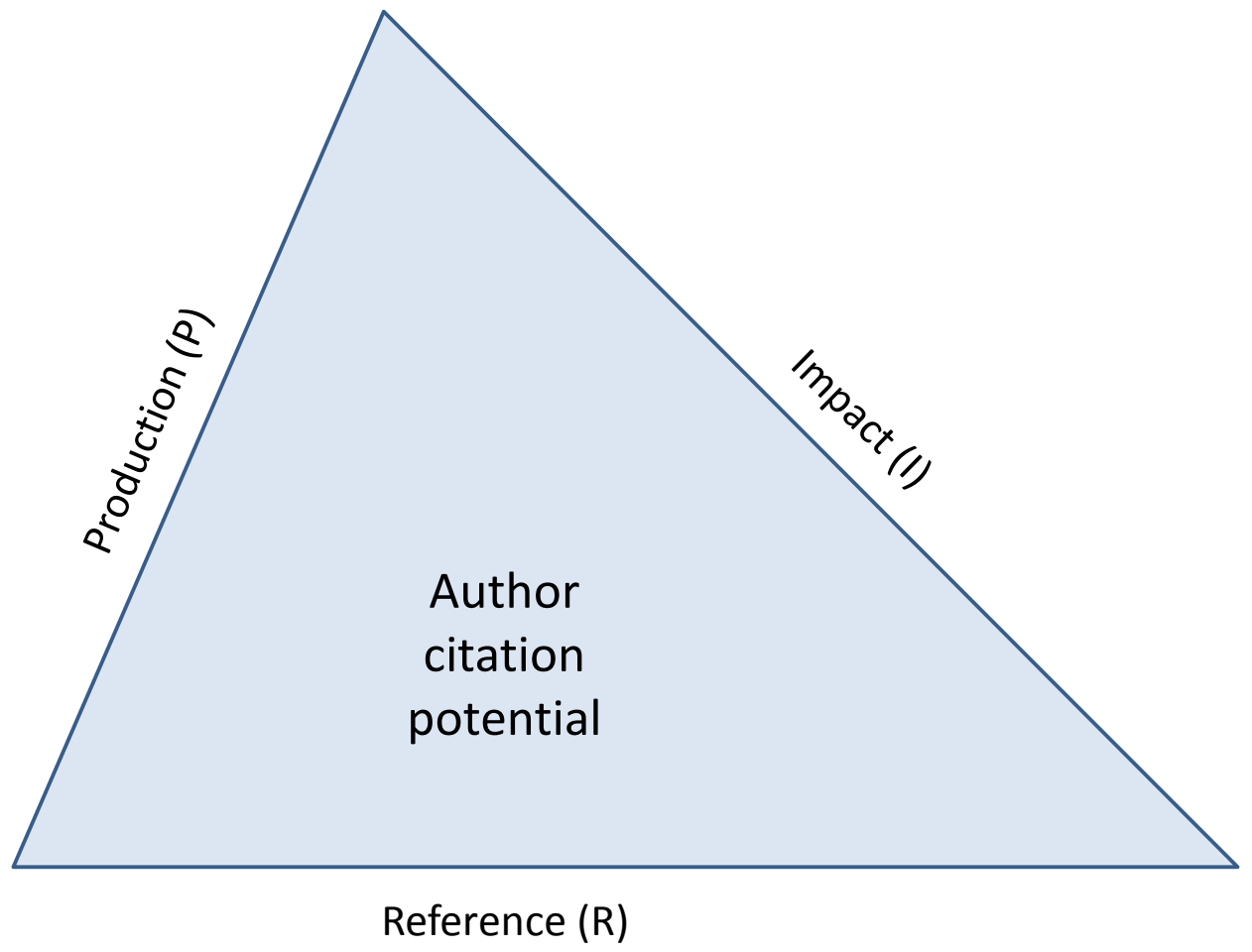

Figure 1: The three dimensions of the author citation potential. The weighted average of the impacts in the journals containing the author's papers ($P$), citing the author's papers ($I$), and included as references in the author's papers ($R$)

Table 1: Operational characterization of the dimensions and proportions of the author citation potential

| Notation | Definition |
|---|---|
| $Y$ | Target window (5-year time window) |
| $JII_y^j$ | Journal impact indicator of $j$ in year $y \in Y$ |
| $j \in J$ | Journals in which the author's papers in target window $Y$ are published |
| $NPub_y^j$ | Number of papers in journal $j$ in year $y$ |
| $NPub_Y^J = \sum_{j \in J} \sum_{y \in Y} NPub_y^j$ | Total number of papers in target window $Y$ |
| $u_y^j = NPub_y^j \,/\, NPub_Y^J$ | Weight of journal $j$ in year $y$ |
| $P_Y = \sum_{j \in J} \sum_{y \in Y} \left( u_y^j \times JII_y^j \right)$ | Production dimension of the author in target window $Y$ |
| $i \in I$ | Journals in which the author's papers in target window $Y$ are cited |
| $NCit_y^i$ | Number of times that year $y$ volumes of journal $i$ cite the author's papers |
| $NCit_Y^I = \sum_{i \in I} \sum_{y \in Y} NCit_y^i$ | Total number of citations in target window $Y$ |
| $v_y^i = NCit_y^i \,/\, NCit_Y^I$ | Weight of journal $i$ in year $y$ |
| $I_Y = \sum_{i \in I} \sum_{y \in Y} \left( v_y^i \times JII_y^i \right)$ | Impact dimension of the author in target window $Y$ |
| $k \in K$ | Journals that author cites in papers in target window $Y$ |
| $NRef_y^k$ | Number of times that year $y$ volumes of journal $k$ are cited by the author's papers |
| $NRef_Y^K = \sum_{k \in K} \sum_{y \in Y} NRef_y^k$ | Total number of references in target window $Y$ |
| $w_y^k = NRef_y^k \,/\, NRef_Y^K$ | Weight of journal $k$ in year $y$ |
| $R_Y = \sum_{k \in K} \sum_{y \in Y} \left( w_y^k \times JII_y^k \right)$ | Reference dimension of the author in target window $Y$ |
| $\left( P / I \right)_Y$ | Production over impact $P_Y \,/\, I_Y$ |
| $\left( P / R \right)_Y$ | Production over reference $P_Y \,/\, R_Y$ |
| $\left( I / R \right)_Y$ | Impact over reference $I_Y \,/\, R_Y$ |
| $\left( \frac{P+I}{2R} \right)_Y$ | Production and impact over reference $\frac{1}{2}\left( \frac{P_Y}{R_Y} + \frac{I_Y}{R_Y} \right) = \frac{P_Y + I_Y}{2R_Y}$ |

Table 2: Production dimension and journals in which Bocci's papers (Physics & Astronomy) are published

| Journals | 2009 | | 2010 | | 2011 | | 2012 | | 2013 | |
|---|---|---|---|---|---|---|---|---|---|---|
| | Papers | SJR | Papers | SJR | Papers | SJR | Papers | SJR | Papers | SJR |
| Physical Review Letters | 25 | 5.264 | 24 | 5.438 | 37 | 5.160 | 32 | 4.537 | 19 | 4.099 |
| Physical Review D | 23 | 2.278 | 16 | 2.176 | 24 | 2.167 | 32 | 2.051 | 28 | 1.899 |
| Physics Letters Section B | 1 | 2.441 | 4 | 2.569 | 16 | 2.566 | 23 | 3.062 | 10 | 4.422 |
| Journal of High Energy Physics | 0 | 1.108 | 2 | 1.232 | 7 | 1.308 | 22 | 0.931 | 14 | 1.027 |
| Journal of Instrumentation | 3 | 0.640 | 22 | 1.430 | 1 | 0.899 | 1 | 0.330 | 1 | 0.840 |
| … | | | | | | | | | | |
| $P_SJR$ = (25×5.264 + 23×2.278 + 1×2.441 + …) / 412 = 2.817 | | | | | | | | | | |

Source: Scopus, 2009-2013; SJR = Scimago journal ranking.

Table 3: Dimensions and proportions of the author citation potential in a set of 120 randomly selected highly productive authors from four subject areas

| | | | | | | SJR | | | | | | | SNIP | | | | | | |
|---|---|---|---|---|---|---|---|---|---|---|---|---|---|---|---|---|---|---|---|
| | | | | | | Dimensions | | | Proportions | | | | Dimensions | | | Proportions | | | |
| | Author | Subject area | # Papers | # Cites | *h* | *P* | *I* | *R* | *P/I* | *P/R* | *I/R* | *(P+I)/2R* | *P* | *I* | *R* | *P/I* | *P/R* | *I/R* | *(P+I)/2R* |
| 1 | Barcelo, D. | Chem | 107 | 1727 | 25 | 1.604 | 1.813 | 2.006 | 0.885 | 0.800 | 0.904 | 0.852 | 1.497 | 1.779 | 1.757 | 0.841 | 0.852 | 1.013 | 0.932 |
| 2 | Alkorta, I. | Chem | 206 | 1468 | 19 | 0.999 | 1.023 | 2.653 | 0.977 | 0.377 | 0.386 | 0.381 | 0.981 | 0.940 | 1.492 | 1.044 | 0.658 | 0.630 | 0.644 |
| 3 | Elguero, J. | Chem | 206 | 1385 | 18 | 0.973 | 1.024 | 2.611 | 0.950 | 0.373 | 0.392 | 0.382 | 0.961 | 0.943 | 1.489 | 1.019 | 0.645 | 0.633 | 0.639 |
| 4 | Fierro, J.L.G. | Chem | 117 | 1056 | 17 | 1.316 | 1.754 | 2.091 | 0.750 | 0.629 | 0.839 | 0.734 | 1.318 | 1.747 | 1.819 | 0.754 | 0.725 | 0.960 | 0.842 |
| 5 | Garcia, H. | Chem | 69 | 1024 | 16 | 2.664 | 2.769 | 6.771 | 0.962 | 0.393 | 0.409 | 0.401 | 1.472 | 1.516 | 3.841 | 0.971 | 0.383 | 0.395 | 0.389 |
| 6 | Jimenez-Barbero, J. | Chem | 85 | 773 | 16 | 2.377 | 3.005 | 3.150 | 0.791 | 0.755 | 0.954 | 0.854 | 1.334 | 1.595 | 1.598 | 0.836 | 0.835 | 0.998 | 0.916 |
| 7 | Arbiol, J. | Chem | 50 | 773 | 16 | 3.860 | 4.230 | 5.460 | 0.913 | 0.707 | 0.775 | 0.741 | 2.100 | 2.103 | 2.948 | 0.999 | 0.712 | 0.713 | 0.713 |
| 8 | Iglesias, M. | Chem | 30 | 567 | 14 | 2.040 | 2.183 | 5.304 | 0.934 | 0.385 | 0.412 | 0.398 | 1.287 | 1.340 | 2.997 | 0.960 | 0.429 | 0.447 | 0.438 |
| 9 | Lopez, F. | Chem | 23 | 544 | 14 | 3.386 | 3.036 | 4.148 | 1.115 | 0.816 | 0.732 | 0.774 | 1.604 | 1.465 | 2.136 | 1.095 | 0.751 | 0.686 | 0.718 |
| 10 | Colmenero, J. | Chem | 66 | 582 | 13 | 1.870 | 1.990 | 1.991 | 0.940 | 0.939 | 0.999 | 0.969 | 1.339 | 1.429 | 1.458 | 0.937 | 0.918 | 0.980 | 0.949 |
| 11 | Veciana, J. | Chem | 49 | 564 | 13 | 2.488 | 2.455 | 5.064 | 1.013 | 0.491 | 0.485 | 0.488 | 1.512 | 1.436 | 2.828 | 1.053 | 0.535 | 0.508 | 0.521 |
| 12 | Tauler, R. | Chem | 53 | 468 | 13 | 1.383 | 1.236 | 1.680 | 1.119 | 0.823 | 0.736 | 0.779 | 1.645 | 1.522 | 1.643 | 1.081 | 1.001 | 0.926 | 0.964 |
| 13 | Fuertes, A.B. | Chem | 22 | 705 | 12 | 2.124 | 1.945 | 3.373 | 1.092 | 0.630 | 0.577 | 0.603 | 1.568 | 1.604 | 2.143 | 0.978 | 0.732 | 0.748 | 0.740 |
| 14 | Molins, E. | Chem | 55 | 432 | 12 | 1.525 | 1.472 | 3.434 | 1.036 | 0.444 | 0.429 | 0.436 | 1.070 | 1.230 | 2.013 | 0.870 | 0.532 | 0.611 | 0.571 |
| 15 | Naffakh, M. | Chem | 24 | 389 | 12 | 2.022 | 1.210 | 1.539 | 1.671 | 1.314 | 0.786 | 1.050 | 1.862 | 1.383 | 1.699 | 1.346 | 1.096 | 0.814 | 0.955 |
| 16 | Herrero, M. | Chem | 25 | 519 | 11 | 1.802 | 1.504 | 1.539 | 1.198 | 1.171 | 0.977 | 1.074 | 1.663 | 1.653 | 1.603 | 1.006 | 1.037 | 1.031 | 1.034 |
| 17 | Lazaro, M.J. | Chem | 46 | 431 | 11 | 1.737 | 1.645 | 1.832 | 1.056 | 0.948 | 0.898 | 0.923 | 1.650 | 1.760 | 1.799 | 0.938 | 0.917 | 0.978 | 0.948 |
| 18 | Sanchez-Cortes, S. | Chem | 44 | 425 | 11 | 1.488 | 1.335 | 1.643 | 1.115 | 0.906 | 0.813 | 0.859 | 1.210 | 1.153 | 1.289 | 1.049 | 0.939 | 0.894 | 0.917 |
| 19 | Kubacka, A. | Chem | 21 | 563 | 10 | 3.594 | 2.560 | 2.487 | 1.404 | 1.445 | 1.029 | 1.237 | 2.325 | 1.784 | 1.726 | 1.303 | 1.347 | 1.034 | 1.190 |
| 20 | Ezquerra, T.A. | Chem | 33 | 296 | 10 | 2.057 | 1.712 | 2.324 | 1.202 | 0.885 | 0.737 | 0.811 | 1.405 | 1.441 | 1.832 | 0.975 | 0.767 | 0.787 | 0.777 |
| 21 | Queralt, I. | Chem | 27 | 223 | 9 | 1.095 | 1.159 | 1.248 | 0.945 | 0.877 | 0.929 | 0.903 | 1.357 | 1.283 | 1.392 | 1.058 | 0.975 | 0.922 | 0.948 |
| 22 | Pons, R. | Chem | 25 | 173 | 9 | 1.538 | 2.034 | 1.551 | 0.756 | 0.992 | 1.311 | 1.152 | 1.427 | 1.629 | 1.383 | 0.876 | 1.032 | 1.178 | 1.105 |
| 23 | Notario, R. | Chem | 40 | 196 | 8 | 1.021 | 1.152 | 2.691 | 0.886 | 0.379 | 0.428 | 0.404 | 1.065 | 1.060 | 1.900 | 1.005 | 0.561 | 0.558 | 0.559 |
| 24 | Roncero, O. | Chem | 27 | 156 | 8 | 1.260 | 1.293 | 1.539 | 0.974 | 0.819 | 0.840 | 0.829 | 1.014 | 1.023 | 1.119 | 0.991 | 0.906 | 0.914 | 0.910 |
| 25 | Torres, J.L. | Chem | 25 | 121 | 8 | 1.292 | 1.276 | 1.518 | 1.013 | 0.851 | 0.841 | 0.846 | 1.391 | 1.604 | 1.545 | 0.867 | 0.900 | 1.038 | 0.969 |
| 26 | Delgado-Barrio, G. | Chem | 27 | 168 | 7 | 1.102 | 1.187 | 1.807 | 0.928 | 0.610 | 0.657 | 0.633 | 0.966 | 0.998 | 1.346 | 0.968 | 0.718 | 0.741 | 0.730 |

| | | | | | | | | | | | | | | | | | | | | |
|---|---|---|---|---|---|---|---|---|---|---|---|---|---|---|---|---|---|---|---|---|
| 27 | Perez-Pariente, J. | Chem | 30 | 165 | 7 | 1.958 | 2.726 | 2.910 | 0.718 | 0.673 | 0.937 | 0.805 | 1.547 | 2.054 | 1.996 | 0.753 | 0.775 | 1.029 | 0.902 |
| 28 | Oro, L.A. | Chem | 20 | 118 | 7 | 1.590 | 1.809 | 3.891 | 0.879 | 0.409 | 0.465 | 0.437 | 1.070 | 1.066 | 2.277 | 1.004 | 0.470 | 0.468 | 0.469 |
| 29 | Canadell, E. | Chem | 25 | 162 | 6 | 2.451 | 1.908 | 4.345 | 1.285 | 0.564 | 0.439 | 0.502 | 1.524 | 1.342 | 2.482 | 1.136 | 0.614 | 0.541 | 0.577 |
| 30 | Senent, M.L. | Chem | 21 | 83 | 5 | 1.074 | 1.236 | 1.290 | 0.869 | 0.833 | 0.958 | 0.895 | 0.999 | 1.044 | 1.050 | 0.957 | 0.951 | 0.994 | 0.973 |
| 31 | Perez, J. | Comp | 32 | 186 | 9 | 1.051 | 1.144 | 1.871 | 0.919 | 0.562 | 0.611 | 0.587 | 1.791 | 1.964 | 2.908 | 0.912 | 0.616 | 0.675 | 0.646 |
| 32 | Torra, V. | Comp | 81 | 271 | 7 | 0.796 | 1.476 | 1.589 | 0.539 | 0.501 | 0.929 | 0.715 | 1.101 | 1.917 | 1.916 | 0.574 | 0.575 | 1.001 | 0.788 |
| 33 | Badia, R.M. | Comp | 50 | 241 | 7 | 0.630 | 0.692 | 1.094 | 0.910 | 0.576 | 0.633 | 0.604 | 1.200 | 1.298 | 2.175 | 0.924 | 0.552 | 0.597 | 0.574 |
| 34 | Labarta, J. | Comp | 25 | 189 | 7 | 0.576 | 0.483 | 1.056 | 1.193 | 0.545 | 0.457 | 0.501 | 1.232 | 0.963 | 2.086 | 1.279 | 0.591 | 0.462 | 0.526 |
| 35 | Rodriguez-Aguilar, J.A. | Comp | 32 | 145 | 6 | 1.028 | 0.902 | 2.349 | 1.140 | 0.438 | 0.384 | 0.411 | 1.956 | 1.486 | 2.526 | 1.316 | 0.774 | 0.588 | 0.681 |
| 36 | Onieva, E. | Comp | 31 | 137 | 6 | 0.815 | 0.939 | 1.591 | 0.868 | 0.512 | 0.590 | 0.551 | 1.424 | 1.716 | 2.598 | 0.830 | 0.548 | 0.661 | 0.604 |
| 37 | Godo, L. | Comp | 44 | 111 | 6 | 0.927 | 1.163 | 1.797 | 0.797 | 0.516 | 0.647 | 0.582 | 1.191 | 1.370 | 2.135 | 0.869 | 0.558 | 0.642 | 0.600 |
| 38 | Bordons, M. | Comp | 11 | 103 | 6 | 1.522 | 1.440 | 1.788 | 1.057 | 0.851 | 0.805 | 0.828 | 1.793 | 1.641 | 2.143 | 1.093 | 0.837 | 0.766 | 0.801 |
| 39 | Villagra, J. | Comp | 27 | 93 | 6 | 0.964 | 1.196 | 1.907 | 0.806 | 0.506 | 0.627 | 0.566 | 1.726 | 2.007 | 2.971 | 0.860 | 0.581 | 0.676 | 0.628 |
| 40 | Milanes, V. | Comp | 39 | 69 | 6 | 0.982 | 1.324 | 1.766 | 0.742 | 0.556 | 0.750 | 0.653 | 1.684 | 2.248 | 2.786 | 0.749 | 0.604 | 0.807 | 0.706 |
| 41 | Haber, R.E. | Comp | 10 | 96 | 5 | 1.052 | 1.611 | 1.724 | 0.653 | 0.610 | 0.934 | 0.772 | 1.843 | 2.019 | 2.604 | 0.913 | 0.708 | 0.775 | 0.742 |
| 42 | Worgotter, F. | Comp | 11 | 61 | 5 | 2.444 | 1.187 | 5.047 | 2.059 | 0.484 | 0.235 | 0.360 | 2.536 | 1.489 | 4.999 | 1.703 | 0.507 | 0.298 | 0.403 |
| 43 | Ortega, J.L. | Comp | 12 | 91 | 4 | 1.382 | 1.158 | 1.375 | 1.193 | 1.005 | 0.842 | 0.924 | 1.617 | 1.372 | 1.790 | 1.179 | 0.903 | 0.766 | 0.835 |
| 44 | Sierra, C. | Comp | 36 | 80 | 4 | 1.070 | 0.859 | 1.354 | 1.246 | 0.790 | 0.634 | 0.712 | 1.950 | 1.366 | 2.093 | 1.428 | 0.932 | 0.653 | 0.792 |
| 45 | Valero, M. | Comp | 30 | 68 | 4 | 1.054 | 0.788 | 1.634 | 1.338 | 0.645 | 0.482 | 0.564 | 1.941 | 1.421 | 2.387 | 1.366 | 0.813 | 0.595 | 0.704 |
| 46 | Arcos, J.L. | Comp | 25 | 65 | 4 | 0.800 | 0.724 | 1.202 | 1.105 | 0.666 | 0.602 | 0.634 | 1.283 | 1.281 | 1.881 | 1.002 | 0.682 | 0.681 | 0.682 |
| 47 | Cazorla, F.J. | Comp | 21 | 49 | 4 | 1.034 | 0.923 | 1.137 | 1.120 | 0.909 | 0.812 | 0.861 | 1.914 | 1.534 | 1.927 | 1.248 | 0.993 | 0.796 | 0.895 |
| 48 | Dellen, B. | Comp | 14 | 47 | 4 | 1.909 | 0.877 | 4.449 | 2.177 | 0.429 | 0.197 | 0.313 | 2.116 | 1.287 | 4.334 | 1.644 | 0.488 | 0.297 | 0.393 |
| 49 | Esteva, M. | Comp | 20 | 44 | 4 | 0.365 | 0.899 | 1.154 | 0.406 | 0.316 | 0.779 | 0.548 | 0.636 | 1.637 | 1.859 | 0.389 | 0.342 | 0.881 | 0.611 |
| 50 | Garcia, E. | Comp | 18 | 40 | 4 | 0.774 | 0.576 | 1.692 | 1.344 | 0.457 | 0.340 | 0.399 | 1.449 | 0.848 | 2.092 | 1.709 | 0.693 | 0.405 | 0.549 |
| 51 | Manya, F. | Comp | 23 | 38 | 4 | 0.655 | 0.529 | 1.328 | 1.238 | 0.493 | 0.398 | 0.446 | 1.083 | 0.794 | 1.767 | 1.364 | 0.613 | 0.449 | 0.531 |
| 52 | Thomas, F. | Comp | 19 | 35 | 4 | 1.252 | 1.176 | 1.376 | 1.065 | 0.910 | 0.855 | 0.882 | 1.825 | 2.027 | 2.295 | 0.900 | 0.795 | 0.883 | 0.839 |
| 53 | Lopez de Mantaras, R. | Comp | 10 | 53 | 3 | 0.738 | 0.648 | 2.936 | 1.139 | 0.251 | 0.221 | 0.236 | 0.972 | 1.125 | 3.393 | 0.864 | 0.286 | 0.332 | 0.309 |
| 54 | Fuster-Sabater, A. | Comp | 28 | 39 | 3 | 0.544 | 0.664 | 1.258 | 0.819 | 0.432 | 0.528 | 0.480 | 0.943 | 1.177 | 1.505 | 0.801 | 0.627 | 0.782 | 0.704 |
| 55 | Quinones, E. | Comp | 12 | 36 | 3 | 0.546 | 0.976 | 1.348 | 0.559 | 0.405 | 0.724 | 0.565 | 1.372 | 1.450 | 2.328 | 0.946 | 0.589 | 0.623 | 0.606 |
| 56 | Plaza, E. | Comp | 24 | 31 | 3 | 0.536 | 0.852 | 1.659 | 0.629 | 0.323 | 0.514 | 0.418 | 0.775 | 1.418 | 2.200 | 0.547 | 0.352 | 0.645 | 0.498 |
| 57 | Unsal, O.S. | Comp | 15 | 21 | 3 | 0.723 | 0.467 | 1.655 | 1.548 | 0.437 | 0.282 | 0.360 | 1.150 | 0.959 | 2.193 | 1.199 | 0.524 | 0.437 | 0.481 |
| 58 | Rius, A. | Comp | 15 | 15 | 3 | 0.233 | 0.454 | 1.576 | 0.513 | 0.148 | 0.288 | 0.218 | 0.325 | 0.710 | 2.209 | 0.458 | 0.147 | 0.321 | 0.234 |
| 59 | Noriega, P. | Comp | 21 | 24 | 2 | 0.978 | 0.915 | 1.194 | 1.069 | 0.819 | 0.766 | 0.793 | 1.488 | 1.459 | 1.984 | 1.020 | 0.750 | 0.735 | 0.743 |
| 60 | Esteve, J. | Comp | 13 | 8 | 1 | 0.236 | 0.901 | 4.018 | 0.262 | 0.059 | 0.224 | 0.141 | 0.371 | 1.303 | 2.519 | 0.285 | 0.147 | 0.517 | 0.332 |
| 61 | Martin, J. | Med | 177 | 2002 | 24 | 2.043 | 1.855 | 4.886 | 1.101 | 0.418 | 0.380 | 0.399 | 1.641 | 1.577 | 2.486 | 1.041 | 0.660 | 0.634 | 0.647 |
| 62 | Gonzalez-Gay, M.A. | Med | 109 | 1238 | 20 | 1.850 | 1.777 | 4.220 | 1.041 | 0.438 | 0.421 | 0.430 | 1.525 | 1.535 | 2.360 | 0.993 | 0.646 | 0.650 | 0.648 |
| 63 | Marcos, A. | Med | 69 | 637 | 13 | 1.044 | 1.366 | 2.410 | 0.764 | 0.433 | 0.567 | 0.500 | 1.171 | 1.272 | 2.102 | 0.921 | 0.557 | 0.605 | 0.581 |
| 64 | Moreno, L.A. | Med | 56 | 526 | 13 | 1.223 | 1.356 | 2.111 | 0.902 | 0.579 | 0.642 | 0.611 | 1.308 | 1.268 | 2.235 | 1.032 | 0.585 | 0.567 | 0.576 |
| 65 | Ortego-Centeno, N. | Med | 52 | 515 | 13 | 2.655 | 1.910 | 5.652 | 1.390 | 0.470 | 0.338 | 0.404 | 1.890 | 1.597 | 2.709 | 1.183 | 0.698 | 0.590 | 0.644 |
| 66 | Alarcon-Riquelme, M.E. | Med | 23 | 438 | 13 | 2.893 | 1.992 | 7.236 | 1.452 | 0.400 | 0.275 | 0.338 | 2.009 | 1.601 | 3.213 | 1.255 | 0.625 | 0.498 | 0.562 |
| 67 | Miranda-Filloy, J.A. | Med | 59 | 543 | 12 | 1.277 | 1.407 | 3.846 | 0.908 | 0.332 | 0.366 | 0.349 | 1.227 | 1.313 | 2.367 | 0.935 | 0.518 | 0.555 | 0.537 |
| 68 | Llorca, J. | Med | 33 | 543 | 12 | 1.201 | 1.518 | 2.490 | 0.791 | 0.482 | 0.610 | 0.546 | 1.198 | 1.362 | 2.029 | 0.880 | 0.590 | 0.671 | 0.631 |
| 69 | Gonzalez-Juanatey, C. | Med | 37 | 537 | 12 | 1.170 | 1.544 | 3.168 | 0.758 | 0.369 | 0.487 | 0.428 | 1.171 | 1.376 | 2.252 | 0.851 | 0.520 | 0.611 | 0.565 |
| 70 | Vazquez-Rodriguez, T.R. | Med | 34 | 482 | 12 | 1.205 | 1.405 | 3.284 | 0.858 | 0.367 | 0.428 | 0.397 | 1.183 | 1.321 | 2.246 | 0.896 | 0.527 | 0.588 | 0.557 |
| 71 | Witte, T. | Med | 34 | 352 | 12 | 2.886 | 1.948 | 5.912 | 1.482 | 0.488 | 0.329 | 0.409 | 2.113 | 1.669 | 3.315 | 1.266 | 0.637 | 0.503 | 0.570 |
| 72 | Esteban, M. | Med | 35 | 350 | 12 | 1.516 | 1.822 | 3.609 | 0.832 | 0.420 | 0.505 | 0.462 | 1.093 | 1.150 | 1.850 | 0.950 | 0.591 | 0.622 | 0.606 |
| 73 | Ruiz, J.R. | Med | 40 | 301 | 11 | 1.123 | 1.270 | 2.087 | 0.884 | 0.538 | 0.609 | 0.573 | 1.306 | 1.242 | 2.069 | 1.052 | 0.631 | 0.600 | 0.616 |
| 74 | Warnberg, J. | Med | 28 | 445 | 10 | 1.448 | 1.592 | 2.484 | 0.910 | 0.583 | 0.641 | 0.612 | 1.408 | 1.507 | 2.064 | 0.934 | 0.682 | 0.730 | 0.706 |
| 75 | Palomino-Morales, R. | Med | 21 | 325 | 10 | 1.458 | 1.445 | 5.326 | 1.009 | 0.274 | 0.271 | 0.273 | 1.304 | 1.322 | 2.909 | 0.986 | 0.448 | 0.454 | 0.451 |
| 76 | Fernandez-Gutierrez, B. | Med | 49 | 312 | 10 | 1.519 | 1.533 | 4.311 | 0.991 | 0.352 | 0.356 | 0.354 | 1.358 | 1.372 | 2.488 | 0.990 | 0.546 | 0.551 | 0.549 |
| 77 | Vyse, T.J. | Med | 20 | 284 | 10 | 3.925 | 2.452 | 7.425 | 1.601 | 0.529 | 0.330 | 0.429 | 2.320 | 1.640 | 3.406 | 1.415 | 0.681 | 0.482 | 0.581 |
| 78 | Covas, M.I. | Med | 21 | 639 | 9 | 1.895 | 2.473 | 3.498 | 0.766 | 0.542 | 0.707 | 0.624 | 1.795 | 2.880 | 4.027 | 0.623 | 0.446 | 0.715 | 0.580 |
| 79 | Urcelay, E. | Med | 25 | 236 | 9 | 1.910 | 1.544 | 8.425 | 1.237 | 0.227 | 0.183 | 0.205 | 1.510 | 1.332 | 3.666 | 1.134 | 0.412 | 0.363 | 0.388 |
| 80 | Balsa, A. | Med | 31 | 224 | 9 | 1.886 | 1.901 | 4.582 | 0.992 | 0.412 | 0.415 | 0.413 | 1.545 | 1.549 | 2.705 | 0.997 | 0.571 | 0.573 | 0.572 |
| 81 | Sanz, Y. | Med | 35 | 166 | 9 | 1.296 | 1.311 | 2.789 | 0.989 | 0.465 | 0.470 | 0.467 | 1.074 | 1.231 | 2.058 | 0.872 | 0.522 | 0.598 | 0.560 |
| 82 | Rodriguez-Rodriguez, L. | Med | 37 | 219 | 8 | 1.616 | 1.516 | 4.216 | 1.066 | 0.383 | 0.360 | 0.371 | 1.440 | 1.325 | 2.475 | 1.087 | 0.582 | 0.535 | 0.559 |
| 83 | Ortega, F.B. | Med | 27 | 196 | 8 | 1.175 | 1.315 | 2.046 | 0.894 | 0.574 | 0.643 | 0.609 | 1.367 | 1.265 | 2.003 | 1.081 | 0.682 | 0.632 | 0.657 |
| 84 | Figuerola, J. | Med | 22 | 139 | 7 | 1.419 | 1.186 | 1.634 | 1.196 | 0.868 | 0.726 | 0.797 | 1.293 | 1.234 | 1.417 | 1.048 | 0.912 | 0.871 | 0.892 |
| 85 | Castaneda, S. | Med | 38 | 141 | 6 | 1.346 | 1.401 | 5.455 | 0.961 | 0.247 | 0.257 | 0.252 | 1.262 | 1.288 | 2.861 | 0.980 | 0.441 | 0.450 | 0.446 |
| 86 | Kafatos, A. | Med | 24 | 110 | 6 | 1.180 | 0.890 | 1.984 | 1.326 | 0.595 | 0.449 | 0.522 | 1.318 | 0.961 | 1.871 | 1.371 | 0.704 | 0.514 | 0.609 |
| 87 | Pons, J.L. | Med | 27 | 100 | 5 | 0.597 | 1.023 | 1.758 | 0.584 | 0.340 | 0.582 | 0.461 | 1.070 | 1.481 | 1.631 | 0.722 | 0.656 | 0.908 | 0.782 |
| 88 | De Henauw, S. | Med | 21 | 83 | 5 | 1.366 | 0.975 | 1.926 | 1.401 | 0.709 | 0.506 | 0.608 | 1.367 | 1.041 | 1.812 | 1.313 | 0.754 | 0.575 | 0.664 |
| 89 | Tobias, A. | Med | 25 | 131 | 4 | 1.054 | 1.519 | 3.009 | 0.694 | 0.350 | 0.505 | 0.428 | 1.239 | 1.743 | 3.384 | 0.711 | 0.366 | 0.515 | 0.441 |
| 90 | Lacasta, C. | Med | 21 | 50 | 4 | 0.273 | 0.604 | 0.898 | 0.452 | 0.304 | 0.673 | 0.488 | 0.350 | 0.842 | 1.216 | 0.416 | 0.288 | 0.692 | 0.490 |
| 91 | Bocci, A. | Phy | 412 | 8780 | 42 | 2.817 | 1.936 | 2.727 | 1.455 | 1.033 | 0.710 | 0.871 | 1.693 | 1.265 | 1.694 | 1.338 | 0.999 | 0.747 | 0.873 |
| 92 | Anastassov, A. | Phy | 372 | 7836 | 39 | 2.754 | 1.913 | 2.808 | 1.440 | 0.981 | 0.681 | 0.831 | 1.675 | 1.264 | 1.719 | 1.325 | 0.974 | 0.735 | 0.855 |
| 93 | Eusebi, R. | Phy | 379 | 7821 | 39 | 2.732 | 1.912 | 2.796 | 1.429 | 0.977 | 0.684 | 0.830 | 1.664 | 1.265 | 1.714 | 1.315 | 0.971 | 0.738 | 0.854 |
| 94 | Vila, I. | Phy | 384 | 7758 | 39 | 2.668 | 1.911 | 2.773 | 1.396 | 0.962 | 0.689 | 0.826 | 1.659 | 1.264 | 1.708 | 1.313 | 0.971 | 0.740 | 0.856 |
| 95 | Giurgiu, G. | Phy | 386 | 7747 | 39 | 2.750 | 1.920 | 2.797 | 1.432 | 0.983 | 0.686 | 0.835 | 1.675 | 1.267 | 1.714 | 1.322 | 0.977 | 0.739 | 0.858 |
| 96 | Nachtman, J. | Phy | 344 | 7500 | 39 | 2.826 | 1.917 | 2.806 | 1.474 | 1.007 | 0.683 | 0.845 | 1.698 | 1.263 | 1.714 | 1.344 | 0.991 | 0.737 | 0.864 |
| 97 | Valuev, V. | Phy | 209 | 5231 | 33 | 2.346 | 1.841 | 2.729 | 1.274 | 0.860 | 0.675 | 0.767 | 1.584 | 1.250 | 1.708 | 1.267 | 0.927 | 0.732 | 0.830 |
| 98 | Paganoni, M. | Phy | 215 | 5125 | 33 | 2.391 | 1.891 | 2.725 | 1.264 | 0.877 | 0.694 | 0.786 | 1.642 | 1.283 | 1.740 | 1.280 | 0.944 | 0.737 | 0.841 |
| 99 | Salerno, R. | Phy | 205 | 5109 | 33 | 2.337 | 1.889 | 2.728 | 1.237 | 0.857 | 0.692 | 0.775 | 1.579 | 1.282 | 1.706 | 1.232 | 0.926 | 0.751 | 0.839 |
| 100 | Fabozzi, F. | Phy | 181 | 4902 | 33 | 2.303 | 1.886 | 2.715 | 1.221 | 0.848 | 0.695 | 0.771 | 1.569 | 1.281 | 1.701 | 1.225 | 0.922 | 0.753 | 0.838 |
| 101 | Jabeen, S. | Phy | 159 | 4658 | 32 | 2.648 | 1.863 | 2.646 | 1.421 | 1.001 | 0.704 | 0.852 | 1.679 | 1.265 | 1.679 | 1.327 | 1.000 | 0.753 | 0.877 |
| 102 | Yamaoka, J. | Phy | 176 | 3616 | 31 | 3.174 | 1.992 | 2.786 | 1.593 | 1.139 | 0.715 | 0.927 | 1.786 | 1.291 | 1.732 | 1.383 | 1.031 | 0.745 | 0.888 |
| 103 | Weinberger, M. | Phy | 72 | 2078 | 27 | 2.144 | 1.843 | 2.573 | 1.163 | 0.833 | 0.716 | 0.775 | 1.522 | 1.263 | 1.659 | 1.205 | 0.917 | 0.761 | 0.839 |
| 104 | Tam, J. | Phy | 46 | 1329 | 22 | 2.838 | 1.861 | 2.748 | 1.525 | 1.033 | 0.677 | 0.855 | 1.765 | 1.256 | 1.737 | 1.405 | 1.016 | 0.723 | 0.870 |
| 105 | Unal, G. | Phy | 66 | 1379 | 20 | 2.529 | 1.941 | 2.564 | 1.303 | 0.986 | 0.757 | 0.872 | 1.558 | 1.304 | 1.694 | 1.195 | 0.920 | 0.770 | 0.845 |
| 106 | Zajacova, Z. | Phy | 55 | 1376 | 20 | 2.683 | 1.943 | 2.582 | 1.381 | 1.039 | 0.753 | 0.896 | 1.614 | 1.305 | 1.709 | 1.237 | 0.944 | 0.764 | 0.854 |
| 107 | Palestini, S. | Phy | 63 | 1371 | 20 | 2.634 | 1.944 | 2.573 | 1.355 | 1.024 | 0.756 | 0.890 | 1.592 | 1.305 | 1.697 | 1.220 | 0.938 | 0.769 | 0.854 |
| 108 | Iakovidis, G. | Phy | 62 | 1368 | 20 | 2.659 | 1.942 | 2.575 | 1.369 | 1.033 | 0.754 | 0.893 | 1.603 | 1.304 | 1.701 | 1.229 | 0.942 | 0.767 | 0.854 |
| 109 | Oakham, F.G. | Phy | 61 | 1341 | 20 | 2.831 | 1.957 | 2.635 | 1.447 | 1.074 | 0.743 | 0.909 | 1.798 | 1.310 | 1.746 | 1.373 | 1.030 | 0.750 | 0.890 |
| 110 | Lagouri, T. | Phy | 60 | 1341 | 20 | 2.622 | 1.949 | 2.563 | 1.345 | 1.023 | 0.760 | 0.892 | 1.591 | 1.307 | 1.692 | 1.217 | 0.940 | 0.772 | 0.856 |
| 111 | Aad, G. | Phy | 61 | 1326 | 20 | 2.653 | 1.949 | 2.568 | 1.361 | 1.033 | 0.759 | 0.896 | 1.605 | 1.307 | 1.696 | 1.228 | 0.946 | 0.771 | 0.858 |
| 112 | Magini, N. | Phy | 87 | 1888 | 19 | 1.857 | 1.815 | 2.539 | 1.023 | 0.731 | 0.715 | 0.723 | 1.391 | 1.258 | 1.643 | 1.106 | 0.847 | 0.766 | 0.806 |
| 113 | Qin, Z. | Phy | 42 | 1249 | 19 | 2.893 | 1.964 | 2.432 | 1.473 | 1.190 | 0.808 | 0.999 | 1.707 | 1.317 | 1.653 | 1.296 | 1.033 | 0.797 | 0.915 |
| 114 | Ebenstein, W.L. | Phy | 48 | 948 | 17 | 2.528 | 2.024 | 2.455 | 1.249 | 1.030 | 0.824 | 0.927 | 1.580 | 1.343 | 1.661 | 1.176 | 0.951 | 0.809 | 0.880 |
| 115 | Xu, G. | Phy | 33 | 863 | 17 | 3.092 | 1.796 | 2.732 | 1.722 | 1.132 | 0.657 | 0.895 | 1.743 | 1.226 | 1.740 | 1.422 | 1.002 | 0.705 | 0.853 |
| 116 | Hackenburg, R. | Phy | 27 | 777 | 15 | 2.825 | 1.772 | 2.662 | 1.594 | 1.061 | 0.666 | 0.863 | 1.678 | 1.221 | 1.723 | 1.374 | 0.974 | 0.709 | 0.841 |
| 117 | D'Ammando, F. | Phy | 36 | 974 | 13 | 3.059 | 2.276 | 3.202 | 1.344 | 0.955 | 0.711 | 0.833 | 1.816 | 1.367 | 2.028 | 1.328 | 0.895 | 0.674 | 0.785 |
| 118 | Caballero, F.G. | Phy | 30 | 112 | 6 | 1.052 | 1.539 | 1.508 | 0.684 | 0.698 | 1.021 | 0.859 | 1.323 | 1.889 | 1.850 | 0.700 | 0.715 | 1.021 | 0.868 |
| 119 | Kadi, Y. | Phy | 25 | 73 | 5 | 0.802 | 0.991 | 2.140 | 0.809 | 0.375 | 0.463 | 0.419 | 0.786 | 1.306 | 1.593 | 0.602 | 0.493 | 0.820 | 0.657 |
| 120 | Ragazzi, S. | Phy | 27 | 46 | 4 | 2.108 | 1.649 | 2.501 | 1.278 | 0.843 | 0.659 | 0.751 | | 1.468 | 1.172 | 1.629 | 1.253 | 0.901 | 0.719 | 0.810 |

Source: Scopus, 2009-2013; Chem = Chemistry; Comp = Computer Science; Med = Medicine; Phy = Physics & Astronomy; SJR = Scimago journal ranking; SNIP = Source normalized impact per paper; *P* = Production; *I* = Impact; *R* = Reference.

Table 4: Central-tendency and variability measures for the dimensions and proportions of the author citation potential in the subject areas

| | | | | | SJR | | | | | | | SNIP | | | | | | |
|---|---|---|---|---|---|---|---|---|---|---|---|---|---|---|---|---|---|---|
| | | | | | Dimensions | | | Proportions | | | | Dimensions | | | Proportions | | | |
| Subject area | Measures | # Papers | # Cites | *h* | *P* | *I* | *R* | *P/I* | *P/R* | *I/R* | *(P+I)/2R* | *P* | *I* | *R* | *P/I* | *P/R* | *I/R* | *(P+I)/2R* |
| Chemistry | Median | 31.5 | 450.0 | 11.5 | 1.671 | 1.733 | 2.406 | 0.968 | 0.777 | 0.720 | 0.792 | 1.398 | 1.439 | 1.742 | 0.984 | 0.771 | 0.826 | 0.872 |
| | Mean | 53.3 | 541.9 | 11.9 | 1.856 | 1.856 | 2.796 | 1.013 | 0.741 | 0.664 | 0.738 | 1.405 | 1.431 | 1.887 | 0.989 | 0.790 | 0.758 | 0.798 |
| | Standard deviation | 48.4 | 424.8 | 4.5 | 0.763 | 0.748 | 1.437 | 0.200 | 0.279 | 0.521 | 0.247 | 0.329 | 0.316 | 0.619 | 0.130 | 0.219 | 0.511 | 0.209 |
| | Min | 20 | 83 | 5 | 0.973 | 1.023 | 1.248 | 0.718 | 0.373 | 0.386 | 0.381 | 0.961 | 0.940 | 1.050 | 0.753 | 0.383 | 0.395 | 0.389 |
| | Max | 206 | 1727 | 25 | 3.860 | 4.230 | 6.771 | 1.671 | 1.445 | 1.311 | 1.237 | 2.325 | 2.103 | 3.841 | 1.346 | 1.347 | 1.178 | 1.190 |
| | Range (Max-Min) | 186 | 1644 | 20 | 2.887 | 3.207 | 5.523 | 0.953 | 1.072 | 0.926 | 0.856 | 1.364 | 1.163 | 2.791 | 0.593 | 0.964 | 0.783 | 0.801 |
| Computer Science | Median | 22.0 | 63.0 | 4.0 | 0.871 | 0.902 | 1.613 | 1.061 | 0.509 | 0.559 | 0.564 | 1.437 | 1.420 | 2.197 | 0.935 | 0.598 | 0.646 | 0.620 |
| | Mean | 25.0 | 82.9 | 4.6 | 0.921 | 0.931 | 1.864 | 1.015 | 0.538 | 0.500 | 0.554 | 1.423 | 1.443 | 2.420 | 1.012 | 0.604 | 0.596 | 0.615 |
| | Standard deviation | 14.9 | 66.2 | 1.7 | 0.462 | 0.307 | 0.984 | 0.428 | 0.220 | 0.312 | 0.203 | 0.520 | 0.388 | 0.736 | 0.370 | 0.207 | 0.527 | 0.167 |
| | Min | 10 | 8 | 1 | 0.233 | 0.454 | 1.056 | 0.262 | 0.059 | 0.197 | 0.141 | 0.325 | 0.710 | 1.505 | 0.285 | 0.147 | 0.297 | 0.234 |
| | Max | 81 | 271 | 9 | 2.444 | 1.611 | 5.047 | 2.177 | 1.005 | 0.934 | 0.924 | 2.536 | 2.248 | 4.999 | 1.709 | 0.993 | 1.001 | 0.895 |
| | Range (Max-Min) | 71 | 263 | 8 | 2.211 | 1.157 | 3.991 | 1.915 | 0.946 | 0.737 | 0.782 | 2.211 | 1.538 | 3.494 | 1.424 | 0.846 | 0.704 | 0.660 |
| Medicine | Median | 33.5 | 318.5 | 10.0 | 1.393 | 1.517 | 3.391 | 0.975 | 0.427 | 0.447 | 0.430 | 1.313 | 1.329 | 2.306 | 0.992 | 0.588 | 0.576 | 0.578 |
| | Mean | 41.0 | 408.8 | 10.3 | 1.582 | 1.528 | 3.756 | 1.008 | 0.450 | 0.407 | 0.459 | 1.395 | 1.410 | 2.441 | 0.998 | 0.583 | 0.578 | 0.589 |
| | Standard deviation | 31.7 | 385.6 | 4.3 | 0.728 | 0.411 | 1.879 | 0.276 | 0.138 | 0.219 | 0.127 | 0.369 | 0.346 | 0.671 | 0.215 | 0.125 | 0.516 | 0.100 |
| | Min | 20 | 50 | 4 | 0.273 | 0.604 | 0.898 | 0.452 | 0.227 | 0.183 | 0.205 | 0.350 | 0.842 | 1.216 | 0.416 | 0.288 | 0.363 | 0.388 |
| | Max | 177 | 2002 | 24 | 3.925 | 2.473 | 8.425 | 1.601 | 0.868 | 0.726 | 0.797 | 2.320 | 2.880 | 4.027 | 1.415 | 0.912 | 0.908 | 0.892 |
| | Range (Max-Min) | 157 | 1952 | 20 | 3.652 | 1.869 | 7.527 | 1.149 | 0.642 | 0.543 | 0.592 | 1.970 | 2.038 | 2.811 | 0.999 | 0.625 | 0.545 | 0.504 |
| Physics & Astronomy | Median | 64.5 | 1377.5 | 20.0 | 2.656 | 1.913 | 2.654 | 1.365 | 0.994 | 0.721 | 0.854 | 1.628 | 1.282 | 1.707 | 1.274 | 0.945 | 0.751 | 0.854 |
| | Mean | 144.1 | 3197.4 | 24.5 | 2.519 | 1.868 | 2.620 | 1.335 | 0.954 | 0.713 | 0.835 | 1.601 | 1.300 | 1.713 | 1.241 | 0.935 | 0.759 | 0.847 |
| | Standard deviation | 133.3 | 2848.3 | 10.9 | 0.523 | 0.205 | 0.274 | 0.212 | 0.157 | 0.749 | 0.099 | 0.190 | 0.117 | 0.074 | 0.177 | 0.104 | 1.578 | 0.044 |
| | Min | 25 | 46 | 4 | 0.802 | 0.991 | 1.508 | 0.684 | 0.375 | 0.463 | 0.419 | 0.786 | 1.172 | 1.593 | 0.602 | 0.493 | 0.674 | 0.657 |
| | Max | 412 | 8780 | 42 | 3.174 | 2.276 | 3.202 | 1.722 | 1.190 | 1.021 | 0.999 | 1.816 | 1.889 | 2.028 | 1.422 | 1.033 | 1.021 | 0.915 |
| | Range (Max-Min) | 387 | 8734 | 38 | 2.372 | 1.285 | 1.694 | 1.038 | 0.815 | 0.557 | 0.580 | 1.030 | 0.717 | 0.435 | 0.820 | 0.539 | 0.347 | 0.258 |

Source: Scopus, 2009-2013; SJR = Scimago journal ranking; SNIP = Source normalized impact per paper; *P* = Production; *I* = Impact; *R* = Reference.

Figure 2: Box-plots comparing the subject areas for the dimensions and proportions of the author citation potential. *P/I* is the ratio based on non-normalized journal impacts that produces the least differences between most areas, which is also close to the results using normalized journal impacts (SNIP)

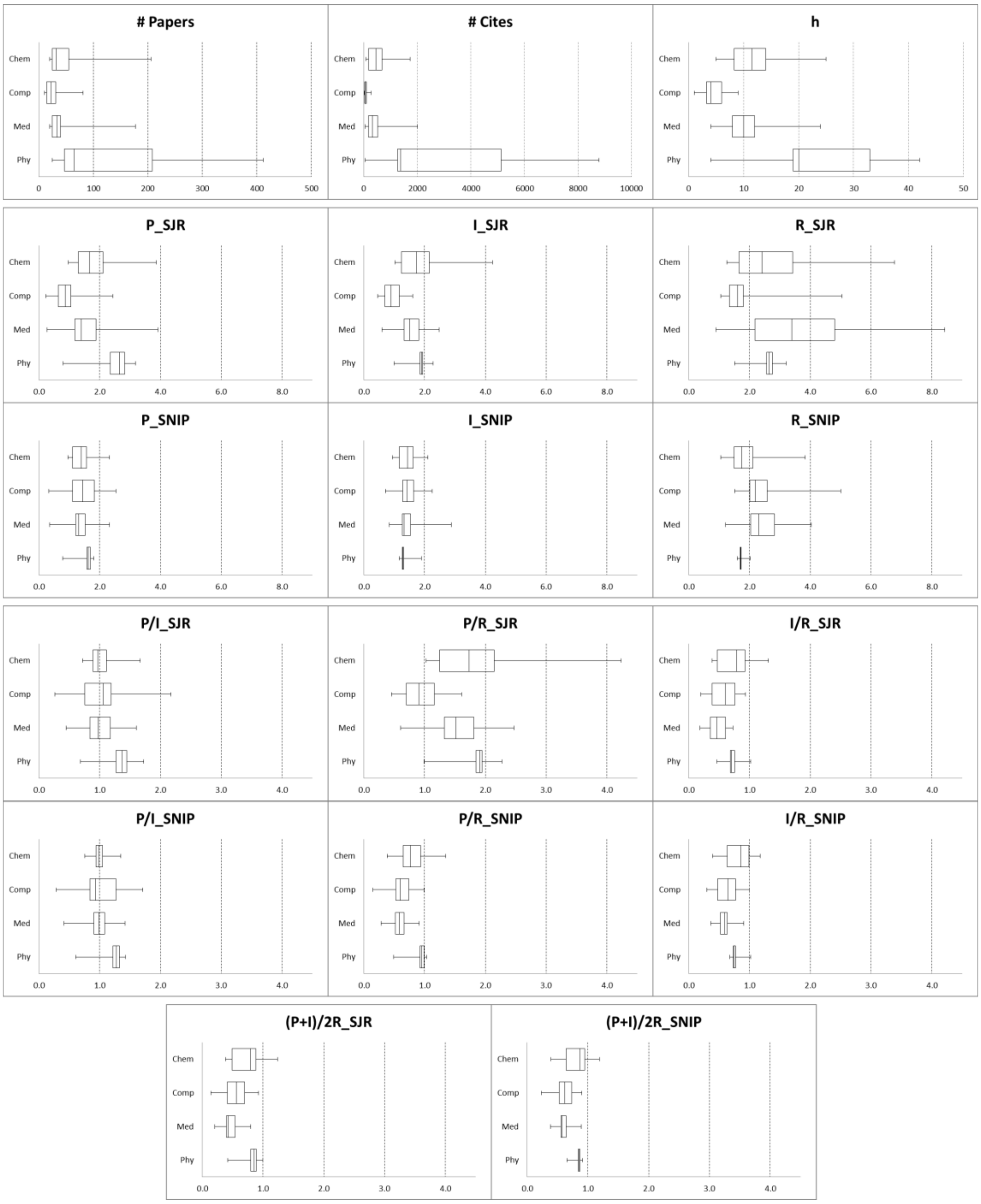

Source: Scopus, 2009-2013; SJR = Scimago journal ranking; SNIP = Source normalized impact per paper; *P* = Production; *I* = Impact; *R* = Reference.

Figure 3: Dimensions for the author citation potential of 120 authors from four subject areas. Impact dimension is quite different from the other dimensions and therefore these three dimensions seem to offer different facets of the author’s research.

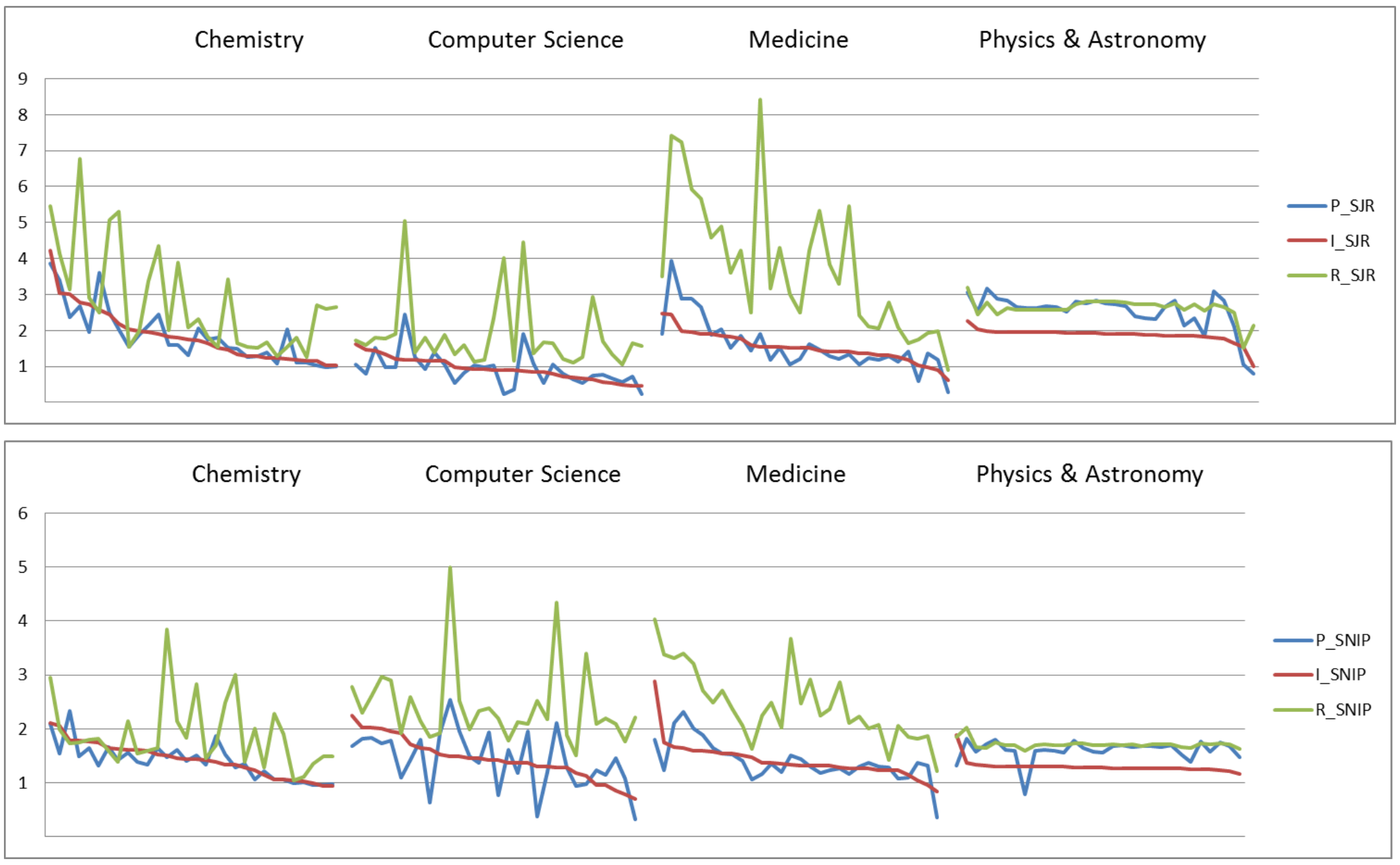

Source: Scopus, 2009-2013; Ordered by *I_SJR* and *I_SNIP*; SJR = Scimago journal ranking; SNIP = Source normalized impact per paper; *P* = Production; *I* = Impact; *R* = Reference.

Table 5: Central-tendency and variability measures for the dimensions and proportions of the author citation potential in the aggregate data

| | SJR | | | | | | | SNIP | | | | | | |
|---|---|---|---|---|---|---|---|---|---|---|---|---|---|---|
| | Dimensions | | | Proportions | | | | Dimensions | | | Proportions | | | |
| Measures | *P* | *I* | *R* | ***P/I*** | *P/R* | *I/R* | *(P+I)/2R* | *P* | *I* | *R* | *P/I* | *P/R* | *I/R* | *(P+I)/2R* |
| Median | 1.521 | 1.526 | 2.564 | **1.065** | 0.610 | 0.714 | 0.629 | 1.493 | 1.319 | 1.908 | **1.042** | 0.710 | 0.658 | 0.710 |
| Mean | 1.719 | 1.546 | 2.759 | **1.093** | 0.671 | 0.696 | 0.647 | 1.456 | 1.396 | 2.115 | **1.060** | 0.728 | 0.623 | 0.712 |
| Range (Max-Min) | 3.692 | 3.776 | 7.527 | 1.915 | 1.386 | 0.881 | 1.096 | 2.211 | 2.170 | 3.949 | 1.424 | 1.200 | 1.128 | 0.956 |
| Within-group variance | 46.360 | 25.089 | 192.557 | 9.972 | 4.924 | 4.037 | 3.711 | | | | 6.705 | | | |
| Between-group variance | 39.434 | 17.325 | 54.463 | 2.358 | 4.547 | 1.455 | 2.639 | | | | 1.321 | | | |
| Percentage reduction of the variance | 14.9% | 30.9% | 71.7% | **76.3%** | 7.7% | 64.0% | 28.9% | | | | **80.3%** | | | |

Source: Scopus, 2009-2013; SJR = Scimago journal ranking; SNIP = Source normalized impact per paper; *P* = Production; *I* = Impact; *R* = Reference; Sd = Standard deviation; Within-group = Within the set of all authors (120); Between-group = Between the subject areas.

Table 6: Pearson correlation coefficients between different indicators of 120 authors

| Subject area | | # Cites | *h* | *P_SJR* | *I_SJR* | *R_SJR* | *P/I_SJR* |
|---|---|---|---|---|---|---|---|
| Chemistry | # Papers | 0.82[c] | 0.72[c] | -0.29 | -0.17 | 0.02 | -0.25 |
| | # Cites | | 0.95[c] | 0.07 | 0.12 | 0.21 | -0.12 |
| | *h* | | | 0.14 | 0.22 | 0.24 | -0.11 |
| Computer Science | # Papers | 0.73[c] | 0.54[b] | -0.17 | 0.20 | -0.29 | -0.26 |
| | # Cites | | 0.85[c] | 0.05 | 0.30 | -0.18 | -0.10 |
| | *h* | | | 0.24 | 0.39 | -0.14 | 0.03 |
| Medicine | # Papers | 0.92[c] | 0.83[c] | 0.08 | 0.16 | 0.09 | 0.00 |
| | # Cites | | 0.92[c] | 0.22 | 0.39 | 0.18 | 0.03 |
| | *h* | | | 0.36 | 0.46[a] | 0.31 | 0.18 |
| Physics & Astronomy | # Papers | 0.99[c] | 0.90[c] | 0.24 | 0.20 | 0.41 | 0.23 |
| | # Cites | | 0.94[c] | 0.28 | 0.24 | 0.46[a] | 0.26 |
| | *h* | | | 0.44[a] | 0.40 | 0.55[b] | 0.40 |

Source: Scopus, 2009-2013; SJR = Scimago journal ranking; *P* = Production; *I* = Impact; *R* = Reference.

[a] significant at the 90% level; [b] significant at the 95% level; [c] significant at the 99% level

Table 7: Spearman rank correlation coefficients between different indicators of 120 authors

| Subject area | | #Cites | *h* | *P_SJR* | *I_SJR* | *R_SJR* | *P/I_SJR* |
|---|---|---|---|---|---|---|---|
| Chemistry | # Papers | 0.69[c] | 0.72[c] | -0.16 | -0.05 | 0.24 | -0.23 |
| | # Cites | | 0.96[c] | 0.25 | 0.26 | 0.41 | -0.01 |
| | *h* | | | 0.19 | 0.21 | 0.36 | -0.02 |
| Computer Science | # Papers | 0.51[b] | 0.53[b] | -0.08 | 0.02 | -0.24 | -0.14 |
| | # Cites | | 0.97[c] | 0.36 | 0.39 | 0.05 | 0.04 |
| | *h* | | | 0.40 | 0.44[a] | 0.09 | 0.05 |
| Medicine | # Papers | 0.57[b] | 0.65[c] | -0.04 | 0.09 | 0.14 | -0.14 |
| | # Cites | | 0.92[c] | 0.30 | 0.57[b] | 0.36 | -0.05 |
| | *h* | | | 0.31 | 0.48[a] | 0.39 | 0.10 |
| Physics & Astronomy | # Papers | 0.98[c] | 0.95[c] | 0.06 | 0.14 | 0.58[c] | 0.11 |
| | # Cites | | 0.97[c] | 0.10 | 0.11 | 0.62[c] | 0.15 |
| | *h* | | | 0.17 | 0.10 | 0.65[c] | 0.24 |

Source: Scopus, 2009-2013; SJR = Scimago journal ranking; *P* = Production; *I* = Impact; *R* = Reference.

[a] significant at the 90% level; [b] significant at the 95% level; [c] significant at the 99% level

Figure 4: Scatter plots between different dimensions of the author citation potential for the 120 authors. These reveal distinct patterns in each subject area in the case of SJR and a more common bivariate distribution across subject areas in the case of SNIP.

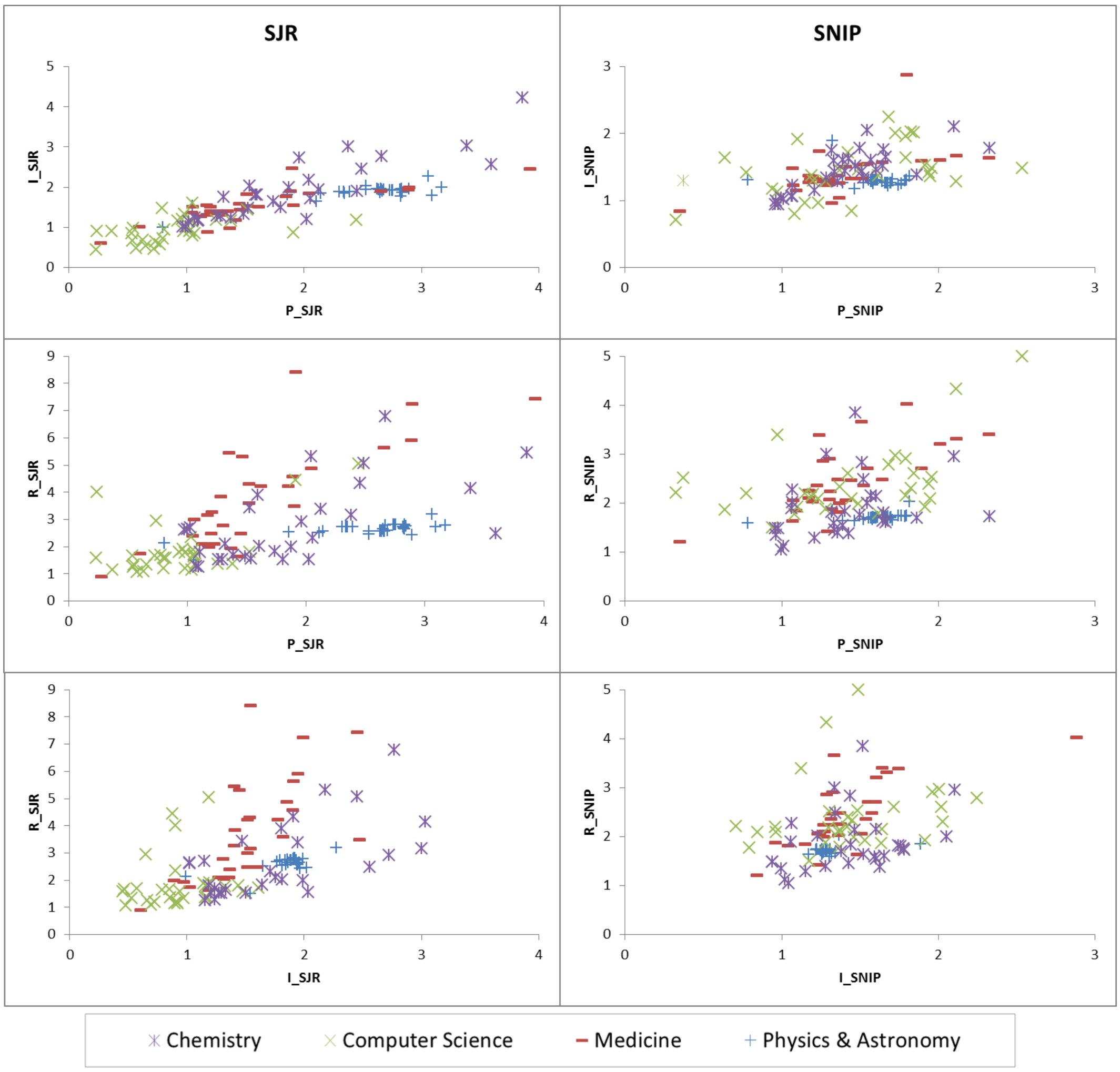

Source: Scopus, 2009-2013; SJR = Scimago journal ranking; SNIP = Source normalized impact per paper; *P* = Production; *I* = Impact; *R* = Reference.